% ------------------------------------------------------------------------- 
% ku.tex (21 June 2002) ( 1 August 2002)
% ------------------------------------------------------------------------- 
\input psfig.sty
% ------------------------------------------------------------------------- 
% Change list publication -> preprint 
% (1) Magnification : either magstep1 and 10pt fonts, or no mag and 12pt 
% (2) Remove \parskip and set \baselineskip = 16pt before Abstract 
% (3) Remove \np before (but not after) references 
% (4) Add \input psfig.sty (or some such) at top 
% Add \psfig commands for each figure 
%-------------------------------------------------------------------------- 
\def\ptitle{Generalized comparison theorems } 
\def\ptitlea{in quantum mechanics} 
% -------------------------------------------------------------------------  
\nopagenumbers 
\magnification=\magstep1 
\hsize 6.0 true in 
\hoffset 0.25 true in % 6 in width with 1.25 in margins default = (6.5, 0) 
\emergencystretch=0.6 in % TEXBook p 107 : allows h-space 
\vfuzz 0.4 in % page-length flexibility 
\hfuzz 0.4 in % line-length flexibility 
\vglue 0.1true in 
\mathsurround=2pt % Default is 2pt 
\def\nl{\noindent} % New line after equations 
\def\nll{\hfil\break\noindent} % Force new line in text 
\def\np{\hfil\vfil\break} % New page 
\def\ppl#1{{\noindent\leftskip 9 cm #1\vskip 0 pt}} % Preprint line 
\def\title#1{\bigskip\noindent\bf #1 ~ \trr\smallskip} % Headings 
 
\def\wbar#1{\overline{#1}} % wide bar
% ------------------------------------------------------------------------- 
% generic unix fonts (all with lower-case names) 
% ------------------------------------------------------------------------- 
\font\trr=cmr10    % Our default 
\font\bf=cmbx10    % Redefinition 
 % small text bold
\font\bmf=cmmib10  % math italic bold 
  % small math italic bold 
 % large math italic bold 
\font\sl=cmsl10    % Redefinition 
 % small text slant
\font\it=cmti10    % Redefinition 
\font\trbig=cmbx10 scaled 1500 % Main Title 
 % Math in Title 
\font\th=cmti10    % Theorems 
\font\tiny=cmr8    % Running title 
% ------------------------------------------------------------------------- 
 % Math Sets eg R -> |R 
\def\mb#1{\hbox{\bmf#1}}    % bold in math mode 
  % small bold in math mode 
          % superscript in text mode  
% -------------------------------------------------------------------------
\def\ng{>\kern -9pt|\kern 9pt} % not greater than 
 % bra < math mode 
 % ket > math mode 
\def\hi#1#2{$#1$\kern -2pt-#2} % hyphen \hi{N}{body} = N-body 
\def\hy#1#2{#1-\kern -2pt$#2$} % hyphen hy{large}{N} = large-N 

\def\sgn{{\rm sgn}}

\def\nl{\noindent}    % new line after equations
\def\nll{\hfil\break} % new line
\def\dbox#1{\hbox{\vrule % Open box size 2#1 (Abrahams p 273) 
\vbox{\hrule \vskip #1\hbox{\hskip #1\vbox{\hsize=#1}\hskip #1}\vskip #1 
\hrule}\vrule}} 
 
\def\qed{\hfill \dbox{0.05true in}} % QED 
 % SQUARE 
% ------------------------------------------------------------------------- 
\output={\shipout\vbox{\makeheadline\ifnum\the\pageno>1 {\hrule} \fi 
{\pagebody}\makefootline}\advancepageno} 
 
\headline{\noindent {\ifnum\the\pageno>1 
{\tiny \ptitle\hfil page~\the\pageno}\fi}} 
\footline{} 
% ------------------------------------------------------------------------- 
\newcount\zz \zz=0 % switch for printing references 
\newcount\q % reference number 
\newcount\qq \qq=0 % starting reference number-1 (usually zero) 
 
\def\pref #1#2#3#4#5{\frenchspacing \global \advance \q by 1 % paper reference 
\edef#1{\the\q}{\ifnum \zz=1 { % 
\item{[\the\q]}{#2} {\bf #3},{ #4.}{~#5}\medskip} \fi}} 
 
\def\bref #1#2#3#4#5{\frenchspacing \global \advance \q by 1 % book reference 
\edef#1{\the\q}{\ifnum \zz=1 { % 
\item{[\the\q]}{#2}, {\it #3} {(#4).}{~#5}\medskip} \fi}} 
 
\def\gref #1#2{\frenchspacing \global \advance \q by 1 % general reference 
\edef#1{\the\q}{\ifnum \zz=1 { % 
\item{[\the\q]}{#2}\medskip} \fi}}

\def\sref #1{~[#1]} % in text line 
 % in text line 
 
\def\references#1{\zz=#1 
\parskip=2pt plus 1pt % default is 0pt plus 1pt 
{\ifnum \zz=1 {\noindent \bf References \medskip} \fi} \q=\qq 

%-----------------------------
%Ref. N dimensions
%-----------------------------
\bref{\som}{A. Sommerfeld}{Partial Differential Equations in Physics}
{Academic, New York, 1949}{The Laplacian in $N$ dimensions is discussed on p. 227}
\bref{\barn}{J. F. Barnes, H. J. Brascamp, and E. H. Lieb}{In: Studies in Mathematical Physics: Essays in Honor of Valentine Bargmann (Edited by E. H. Lieb, B. Simon, and A. S. Wightman)}{Princeton University Press, Princeton, 1976}{p 83}
\pref{\andr}{K. Andrew and J. Supplee, Am. J. Phys.}{58}{1177 (1990) }{}
\bref{\movr}{H. Movromatis}{Exercises in Quantum Mechanics}{Kluwer, Dordrecht, 1991}{}
\pref{\hallpn}{R. L. Hall and N. Saad, J. Math. Phys.}{38}{4904 (1997)}{}
\pref{\hallkr}{R. L. Hall and N. Saad, J. Chem. Phys.}{109}{2983 (1998) }{}
\pref{\burg}{F. Burgbacher, C. L\"ammerzahl and A. Macias, J. Math. Phys.}{40}{625 (1999)}{}
\pref{\nieto}{J. Negro, L. M. Nieto and O. Rosas-Ortiz, J. Math. Phys.}{41}{7964 (2000)}{}
\pref{\schl}{W. P. Schleich and J. P. Dahl, Phys. Rev. A}{65}{052109 (2002)}{}
% ------------------------------------------------------------ 

\bref{\reed}{M. Reed and B. Simon}{Methods of Modern Mathematical Physics IV: Analysis of Operators}{Academic, New York, 1978}{The min-max principle for the discrete spectrum is discussed on p75}
\pref{\narn}{H. Narnhoffer and W. Thirring, Acta Phys. Austriaca}{41}{281 (1975)}{}
\bref{\thir}{W. Thirring}{A Course in Mathematical Physics 3: Quantum Mechanics of Atoms and Molecules}{Springer, New York, 1981}{The min-max principle for the discrete spectrum is discussed on p152; the concavity of $E(v)$ is discussed on p 154.}
\pref{\hallkp}{R. L. Hall, J. Math. Phys.}{25}{2078 (1984)}{}
\pref{\hallpow}{R. L. Hall, Phys. Rev. A}{39}{5500 (1989)}{}
\pref{\hallenv}{R. L. Hall, J. Math. Phys.}{94}{2779 (1993)}{}
\bref{\gelfand}{I. M. Gel'fand and S. V. Fomin}{ Calculus of Variations}{Prentic-Hall, Englewood Cliffs, NJ, 1963}{Legendre transformations are discussed on p71.}
\pref{\halli}{R. L. Hall, Phys. Rev. A}{50}{2876 (1994)}{} %flat bottoms
\pref{\hallw}{R. L. Hall, W. Lucha, and F. F. Sch\"oberl, Int. J. Mod. Phys. A}{17}{1931 (2002)}{}
\pref{\hallt}{R.L. Hall, J. Phys. A}{25}{4459 1992}{}
\pref{\rhallcp}{ R.L Hall, Phys. Rev. D}{30 }{433 (1984)  }{ }
\pref{\rhallsum}{R.L Hall, J. Math. Phys.}{33}{1710 (1991) }{}
\pref{\hallprd}{R. L. Hall, Phys. Rev. D}{37}{540 (1988)}{}
\pref{\hallmix}{R. L. Hall, J. Math. Phys.}{33}{1710 (1992)}{}
\pref{\weyl}{H. Weyl, Math. Ann.}{71}{441 (1911)}{}
\pref{\fan}{Ky Fan, Proc. Nat. Acad. Sci. (U.S.)}{35}{652 (1949)}{}
\bref{\wein}{A. Weinstein and W. Stenger}{Methods of Intermediate Problems for
 Eigenvalues}{Academic, New York, 1972}{Weyl's theorem is discussed on p. 163.} 
\bref{\corant}{R. Courant and F. John}{Introduction to Calculus and Analysis II}{A Wiley-Interscience Publication, New York, 1974}{}
\pref{\varma} {S. N Biswas, K. Datt, R. P. Saxena, P. K. Strivastava, and V. S. Varma, J. Math. Phys., No. 9}{14}{1190 (1972)}{}
\pref{\castro }{Francisco M., Ferndez and Eduardo A. Castro, Am. J. Phys., No. 10}{50}{921 (1982)}{}
\pref{\hio} {F. T. Hioe, Don MacMillen, and E. W. Montroll, J. of Math. Phys., No 7}{17}{(1976)}{}
\pref{\trus}{H. Turschner, J.Phys. A, No. 4}{12}{451 (1978)}{}
\pref{\hill}{B. J. B. Crowley and T. F. Hill, J. Phs. A, No. 9}{12}{223 (1979)}{}
\pref{\mark}{Mark S. Ashbaugh and John D. Morgan III, J. Phys. A}{14}{809 (1981)}{}
\pref{\reno}{R. E. Carndall and Mary Hall Reno, J. Math. Phys. }{23}{ 64 (1982) }{}
\pref{\rhalla}{R. L. Hall, J. Math. Phys.} {24} {324 (1983)}{}
\pref{\rhallb}{R.L Hall, J. Math. Phys. }{25} {2708 (1984)}{}
\pref{\hallns}{R. L. Hall and N. Saad, Phys. Lett. A}{237}{107 (1998)}{}
% ------------------------------------------------------------ 
} 
 
\references{0} % Initialization of reference numbers 
% --------------------------------------- end our ref.tex ----------------- 
\topskip 2pt
% ------------------------------------------------------------------------- 
% Title page and Abstract 
% -------------------------------------------------------------------------  
\trr % our standard font 
% ------------------------------------------------------------------------- 
% preprint data using \ppl
% ------------------------------------------------------------------------- 
% preprint data using \ppl 
% ------------------------------------------------------------------------- 
\ppl{CUQM-93}
\ppl{math-ph/0208047} 
\ppl{August 2002}\medskip 
% ------------------------------------------------------------------------- 

% spacing: remove next line for preprint version
%\parskip=5pt plus 1pt      % MAIN PARSKIP remove for preprint
%\baselineskip = 18true pt  % baselineskip paper 18 preprint 16
% ---------------------------------------------------------------------
%\centerline{\bf Abstract}\medskip
% ---------------------------------------------------------------------

%-----------------------------
\vskip 0.4 true in 
\centerline{\trbig \ptitle}
\vskip 0.2 true in
\centerline{\trbig \ptitlea}
\vskip 0.5true in
\baselineskip 12 true pt % for address only                                     
\centerline{\bf Richard L. Hall and Qutaibeh D. Katatbeh}\medskip
\centerline{\sl Department of Mathematics and Statistics,}
\centerline{\sl Concordia University,}
\centerline{\sl 1455 de Maisonneuve Boulevard West,}
\centerline{\sl Montr\'eal, Qu\'ebec, Canada H3G 1M8.}
\vskip 0.2 true in
\centerline{email:\sl~~rhall@cicma.concordia.ca}
\bigskip\bigskip
% ------------------------------------------------------------------------- 
% spacing: remove next line for preprint version 
%\parskip=5pt plus 1pt % MAIN PARSKIP remove for preprint 
\baselineskip = 18true pt % baselineskip paper 18 preprint 16 
 
% ------------------------------------------------------------------------- 
\centerline{\bf Abstract}\medskip 
%-------------------------------------------------------------------------  
This paper is concerned with the discrete spectra of Schr\"odinger operators $H = -\Delta + V,$
where $V(r)$ is an attractive potential in $N$ spatial dimensions. Two principal 
results are reported for the bottom of the spectrum of $H$ in each angular-momentum subspace ${\cal H}_{\ell}$: (i) an optimized
lower bound when the potential is a sum of terms $V(r) = V^{(1)}(r) + V^{(2)}(r)$, and the 
bottoms of the spectra of  $-\Delta + V^{(1)}(r)$ and  $-\Delta + V^{(2)}(r)$ in ${\cal H}_{\ell}$ are known, and (ii) a
 generalized comparison theorem which predicts spectral ordering when the graphs of the comparison potentials $V^{(1)}(r)$ and $V^{(2)}(r)$ intersect in a controlled way.  Pure power-law potentials are studied and an application of the results to the Coulomb-plus-linear potential $V(r) = -a/r + br$ is 
presented in detail: for this problem an earlier formula for energy bounds is sharpened and generalized to $N$ dimensions.
 

\topskip 20pt
\np
% ------------------------------------------------------------------------- 
\title{1.~~Introduction} 
% -------------------------------------------------------------------------
This paper has two principal aspects: the potential-sum approximation, and the generalization of the comparison theorem of quantum mechanics to cases where the comparison potentials intersect. We study spherically-symmetric problems in $N$ spatial dimensions.  There is much interest in problems posed in arbitrary dimension $N$ \sref{\som--\schl}, rather than specifically, say, for $N = 1,$ or $N = 3.$  References\sref{\som} and \sref{\movr} are useful for technical results such as the form of the Laplacian in \hi{N}{dimensional} spherical coordinates; the other papers are concerned with solving problems such as the hydrogen atom\sref{\andr, \burg} and the linear, harmonic-oscillator, hydrogen atom, and Morse potentials\sref{\nieto} in higher dimensions than $N = 3.$  The geometrical methods we use in this paper are independent of dimension, which can usually be carried as a free parameter $N.$  We consider examples with Hamiltonians of the form, $H = -\Delta + v\ \sgn(q)r^q$, or with sums of such potential terms. We suppose that the  Hamiltonian operators $H = -\Delta + V(r),$ $r = \|\mb{r}\|,$ have domains ${\cal D}(H)\subset L^2(R^N),$ they are bounded below, essentially self adjoint, and have at least one discrete eigenvalue at the bottom of the spectrum.  Because the potentials are spherically symmetric,
the discrete eigenvalues $E_{n\ell}$ can be labelled by two quantum numbers, the total angular momentum $\ell = 0,1,2,\dots,$ and a `radial' quantum number, $n = 1,2,3,\dots,$ which counts the eigenvalues in each angular-momentum subspace.  Since the discrete spectrum may be characterized variationally\sref{\reed,\thir}, the elementary {\bf comparison theorem} $V^{(1)} < V^{(2)}
 \Rightarrow E^{(1)}_{n\ell} < E^{(2)}_{n\ell}$ immediately follows. The generalization we shall study (in Section~3) involves comparison potentials whose graphs `cross over' in such a way that spectral ordering is still guaranteed.

Before we study the generalized comparison theorem, we shall need some established results concerning `kinetic potentials'\sref{\hallkp} and `envelope theory'\sref{\hallpow, \hallenv}.  In order to fix ideas and simplify the presentation, let
us suppose that $E$ is a discrete eigenvalue at the bottom of the spectrum of $H = -\Delta + V$ in $N$ dimensions.  It follows that $E = \inf(\psi,H\psi)$ where $\psi\in {\cal D}(H),$ and $\|\psi\| = 1.$  We perform the total minimization in two stages: first we constrain the process by fixing the mean kinetic energy $(\psi,-\Delta\psi) = s,$ and then we minimize over $s > 0.$ The mean potential-energy function under the constraint is called the `kinetic potential' $\wbar{V}(s)$ 
associated with the potential $V(r).$ Thus we define
$$\wbar{V}(s) = \inf_{{{\scriptstyle \psi \in {\cal D}(H)} \atop {\scriptstyle (\psi,\psi) = 1}} \atop {\scriptstyle (\psi, -\Delta\psi) = s}} (\psi, V\psi)\quad \Rightarrow\quad E = \min_{s>0}\left\{s + \wbar{V}(s)\right\}.\eqno{(1.1)}$$
\nl The variational definition of the kinetic potentials implies that (i) $\wbar{cV}(s) = c\wbar{V}(s),$ and (ii) $\wbar{V}^{(1)}(s) \leq \wbar{V}^{(2)}(s)\Rightarrow E^{(1)} \leq E^{(2)}.$ Kinetic potentials can be defined\sref{\hallkp} for higher eigenvalues and they can then be reconstructed from `energy trajectories', the functions which describe how the eigenvalues vary with the coupling parameter $v>0.$  We have in general for coupling
$$H = -\Delta + vf(r)\rightarrow E_{n\ell} = F_{n\ell}(v)\eqno{(1.2a)}$$
\nl and 
$$s = F_{n\ell}(v) - vF_{n\ell}'(v),\quad \wbar{f}_{n\ell}(s) = F_{n\ell}'(v),\eqno{(1.2b)}$$
\nl where $E_{n\ell}$ is the $n$th eigenvalue in the angular-momentum space labelled by $\ell,$
and $F_{n\ell}(v)$ describes how this eigenvalue depends on the coupling $v > 0;$ the corresponding kinetic potentials may then be defined by (1.2b). The relationship $F(v)\leftrightarrow \wbar{f}(s)$ is essentially a Legendre transformation\sref{\gelfand}: for the ground state (or the bottom $E_{1\ell}$ of each angular-momentum subspace) $F(v)$ is concave\sref{\narn, \thir, \hallkp} and consequently $\bar{f}(s)$ is convex: it follows\sref{\halli} immediately from (1.2b) that $F''(v)\bar{f}''(s) = -v^{-3} < 0$ whenever $F''(v) \neq 0;$ thus in general $F(v)$ and $\bar{f}(s)$ have opposite convexities almost everywhere.  For the important class of examples $H = -\Delta + v\ \sgn(q)r^q,$ corresponding to pure powers $q > -2,$ we know that $F(v)$ is {\it concave} for every (discrete) eigenvalue since, by scaling arguments, we have $F_{n\ell}(v) = F_{n\ell}(1) v^{2\over{2+q}},$ and $\sgn(F_{n\ell}(1)) = \sgn(q).$
 
\par The main purpose for this two-step reformulation of `min-max' is that certain spectral approximations are very effectively developed in terms of kinetic potentials.  We shall consider first the `envelope approximation', which in its most succinct form can be summarized as follows
$$f(r) = g(h(r))\quad\Rightarrow\quad \wbar{f_{n\ell}}(s)\approx g(\wbar{h_{n\ell}}(s)).\eqno{(1.3)}$$
\nl Here $f(r)$ is a smooth transformation of a `base' potential $h(r).$ We suppose that the  transformation $g$ is monotone increasing and, if it also has definite convexity, the following important conclusions may be drawn: if $g$ is concave, we get an upper bound $\approx = \leq;$
and, if $g$ is convex, we obtain a lower bound $\approx = \geq.$  These results may also be derived by the use families of upper and lower `tangential' potentials\sref{\hallw}. In Section~2 we shall apply this
result to study the Coulomb-plus-linear potential $V(r) = -1/r + \lambda r$ which is clearly at once a convex transformation of the Hydrogenic potential $h(r) = -1/r$ and a concave transformation of the linear potential $h(r) = r.$  We shall show that we are also able to express both the upper and lower bounds for the entire discrete spectrum in the form of explicit rational functions $\lambda=\lambda(E_{n\ell}).$

The base potentials used for the Coulomb-plus-linear potential are both pure powers.  Thus we shall need to use the corresponding base kinetic potentials.  In fact we have shown in general\sref{\hallpow} that 
$$-\Delta + \sgn(q)r^q \quad\Rightarrow \quad E_{n\ell}^{N} = \min_{r>0}\left\{\left({{P^N_{n\ell}(q)}\over{r}}\right)^{2} + \sgn(q)r^q\right\},\eqno{(1.4)}$$
\nl where, for example, $P^N_{n\ell}(-1) = (n+\ell+N/2-3/2),$ and $P^N_{n\ell}(2) = (2n+\ell+N/2-2).$  These \hi{P}{numbers} and the underlying eigenvalues $E_{n\ell}^N$ 
satisfy the relation $E_{n\ell}^N = E_{n 0}^{N+2\ell}:$ this is generally true for central potentials and is the content of Theorem~2, which we prove in Section~4.  Numerical values for $P^N_{n 0}(1)$ are given in Table~1 for $N=2,...,12$. It is interesting that the case $q = 0$ corresponds {\it exactly} to the $\ln(r)$ potential\sref{\hallenv}.  The expression in (1.4) is derived by a change of variable $s\rightarrow (P^N/r)^2$ in the kinetic-potential formalism.  The application to the Coulomb-plus-linear potential is not our only interest in these \hi{P^N}{numbers}.  They provide through (1.4) 
a nice representation for the pure-power eigenvalues since the \hi{P^N}{numbers} vary smoothly with
$q$ through $q = 0$ whereas the eigenvalues themselves do not\sref{\hallpow}.  We have proved\sref{\hallpow} that $P^N_{n\ell}(q)$ are monotone increasing in $q$.  This result was obtained by using envelope theory: we considered one power $q$ as a smooth transformation of another $p$, and then took the limit $p\rightarrow q.$  

In Section~3 we prove Theorem~1 which provides a lower bound for the bottom of the spectrum in each angular momentum subspace using the sum approximation. In Section~4 we prove Theorem 2, which establishes the invariance of the eigenvalues with respect to changes in $\ell$ and $N$ that leave the sum $N + 2\ell$ invariant. This allows us to restrict our considerations to the ground state in sufficiently high dimension $N.$  We reformulate the {\bf refined comparison theorem} (Theorem~3 of Ref.\sref{\hallt}) which becomes Theorem~3 here. We first prove the monotonicity of the ground-state wave function in $N$ dimensions; then we prove Theorem~4, which extends Theorem~3 to $N\ge 2$ dimensions.  Finally we prove Theorems~5, 6, and 7 which provide simple explicit sufficient conditions for the application of Theorem~4 under a variety of crossing schemes.  In Section~4 we apply Theorem~5 to sharpen the envelope bounds already found in Section~2 for the bottom of the spectrum $E$ of $H$ when $V$ is the Coulomb-plus-linear potential $V(r) = -a/r + b r.$

%---------------------------------------------------------------
\title{2.~~Coulomb-plus-linear potential: an eigenvalue formula }
%---------------------------------------------------------------
 The Coulomb-plus-linear potential $V(r) = -a/r + br$ is of interest in physics because it serves  as a nonrelativistic model for the principal part of the quark-quark interaction. 
First, we will use the envelope method to derive a simple formula for upper and lower bounds for all the eigenvalues $E_{n\ell},~n=1,2,3,...,~\ell=0,1,2,\dots$ Because the linear potential, rather than the harmonic oscillator, is used as a basis for the upper bound, the new bounds are sharper than those of the earlier paper\sref{\rhallcp}.

   If we denote the eigenvalues of $H=-\omega\Delta-\alpha/r +\beta r$ by $E(\omega,\alpha,\beta)$ and consider a scale of change of the form
$r'=r/\sigma$, and if we further choose $\sigma=\alpha/\omega$, then it is easy to show that
$$E(\omega,\alpha,\beta)=\alpha^{2} {\omega^{-1}}E(1,1,\lambda),~\lambda={\beta\omega^2\over\alpha^3}. \eqno{(2.1)} $$
\noindent Thus it is sufficient to study the special case $ H=-\Delta -1/r+\lambda r.$

 We need a solvable model which we can use as an envelope basis. The natural bases to use in the present context are the hydrogenic and linear potentials
$$  h(r)=\sgn(q)r^q,\quad{\rm where}\quad q=-1,1.  \eqno{(2.2)} $$
\noindent The spectrum generated by the potential $h(r)$ is represented precisely by means of the 
 semi-classical expression $(1.1)$ as follows: 
$${\cal E}_{n\ell}(v) = \min_{s > 0}\{s + v\bar{h}_{n\ell}(s)\},\eqno{(2.3)}$$
where the `kinetic potentials' $\bar{h}_{n\ell}(s)$ associated with the  power-law potentials (1.1) are given\sref{\hallenv} by
$$\bar{h}_{n\ell}(s)={2\over q}\left |{q{\cal E}_{n\ell}^{(q)}\over {2+q}}\right|^{{q+2\over 2}}s^{{-q/ 2}},\eqno{(2.4)} $$
and ${\cal E}_{n\ell}^{(q)}$ is the eigenvalue of $-\Delta+\sgn (q)r^q $ in $N$ dimensions, that is to say, corresponding to the pure-power potential with coupling 1. If we use the potential $h(r)=-{1\over r}$ as an envelope basis, then $V(r)=-{1\over r}+\lambda r=g(-{1\over r}) $ implies $g$ is convex. And if we use the linear potential $h(r)=r$ as an envelope basis, then $g$ is concave. A weaker upper bound is provided by the harmonic oscillator $h(r)=r^2,$ for which, again $g(h)$ is convex.

For the power-law potentials $h(r)=\sgn(q)r^q$ we can simplify (2.3) by changing the minimization variable $s$ to $r$ defined in each case by the equation $\bar{h}_{n\ell}(s)=h(r)$ so that $g(h(r))=f(r)={-{1\over r}+{\lambda r}}$. The minimization on the other hand, which yields eigenvalue approximations
 for the Hamiltonian $H = -\omega\Delta + f(r)~(\omega > 0),$  can be expressed in the form
 $$E_{n\ell}^N\approx\min_{r>0}\left\{\omega \left({P_{n\ell}^N(q)\over r}\right)^2-{1\over r}+\lambda r \right\}, \eqno{(2.5)} $$
where 
$$P^N_{n\ell}(q)=\left |E_{n\ell}^{(q)}\right|^{{2+q\over 2q}}\left[{2\over {2+q}}\right]^{{1\over q}}\left |{q\over {2+q}}\right |^{1\over 2},~q\ne 0. \eqno{(2.6)} $$

\noindent We obtain a lower bound with $P^N_{n\ell}(-1)=(n+\ell+N/2-3/2)$
and the harmonic-oscillator upper bound (of Ref.[\rhallcp]) with  $P_{n\ell}^N(2)=2n+\ell+N/2-2,$ and a sharper upper bound with  
$P^N_{n\ell}(1);$ the \hi{P^N_{n 0}(1)}{numbers} are provided in Table 1 for $N=2,...,12.$ This table allows $\ell > 0$ since $P^N_{n\ell} = P^{N+2\ell}_{n 0}:$ it is clear that $E_{n\ell}^N(-1)$ and $E_{n\ell}^N(2),$ and the corresponding P-numbers, are invariant with respect to changes in $\ell$ and $N$ which preserve the sum $2\ell+N;$ this symmetry is also true for $E_{n\ell}^N(1),$ indeed for {\it all} eigenvalues generated by a central potential. This property is the content of Theorem~2, which we state and prove is Section~4.
\noindent We thus obtain the following energy bounds 

$$ \min_{r>0}\left\{\left({P^N_{n\ell}(-1)\over r}\right)^2-{1\over r}+\lambda r)\right\}  \le E_{n\ell} \le \min_{r>0}\left\{\left({P^N_{n\ell}(1)\over r}\right)^2-{1\over r}+\lambda r\right\} \eqno{(2.7)}  $$

\noindent for $n=1,2,3,...,~\ell =0,1,2,....$ Consequently, the energy bounds are given by the parametric equations 
$$E_{n\ell}=-{1\over {2\nu t}}+{3\lambda \mu t\over 2} \eqno{(2.8{\rm a})} $$
$$ 1={t\over {2\nu}}+{\lambda \mu t^3\over 2},\quad t=rP^N_{n\ell}(q)\quad q=-1,1,\eqno{(2.8{\rm b})} $$

\noindent wherein the lower and upper bounds take the values $\nu=\mu=P^N_{n\ell}(-1)$ and $ \nu=\mu=P^N_{n\ell}(1)$ respectively. It is interesting that we can actually solve Eqs. (2.8a) and (2.8b) to obtain $\lambda$ as an explicit function of $E=E^N_{n\ell};$ the result namely is 
$$\lambda={\left\{ 2(\nu E)^3-\nu E^2\left[(1+3\nu^2 E)^{1\over 2}-1\right]\right \}\over \mu\left[(1+3 \nu^2E)^{1\over 2}-1\right ]^3}\eqno{(2.9)} $$

\noindent with $E\ge -{1\over {4\nu^2}}$ (corresponding to $\lambda=0$ for the pure hydrogenic spectrum). We emphasis that these bounds are valid for all the discrete eigenvalues in arbitrary dimension $N\ge 2.$ The bounds are weak for  $n>1$, but at the bottom of each angular momentum subspace $n=1$ they are sharp and improve with increasing $\ell,N,$ and $\lambda.$ The lower bound for the bottom of each angular-momentum subspace $(n=1)$ can be improved by use of the `sum approximation' (\sref{\rhallsum} and Section~3 below) in which  $\nu=P^N_{n\ell}(1)$ (Table~1) and $\mu=P^N_{n\ell}(-1)=(n+\ell+N/2-3/2)$. In Figure~1 we exhibit these bounds for $n=1$, $N=3,$ and $\ell=0,1,2,3.$
% ------------------------------------------------------------------------- 
\title{3.~~The sum approximation: lower bounds} 
% -------------------------------------------------------------------------

We now consider potential which are sums of terms. 
Since further generalizations easily follow, we first look at the problem of the sum of only two potential terms. We assume that each potential $vh^{(i)}(r)$ alone, when added to the kinetic-energy operator $-\Delta,$ has a discrete eigenvalue $E$ at the bottom of the spectrum for sufficiently large coupling $v.$  We note that the proof is unchanged if we restrict the problem to a given angular-momentum subspace labelled by $\ell;$ our claim then concerns the
bottom of the spectrum of $H$ in such a subspace; in the more general case, all the kinetic potentials would be labelled by $\ell.$ We express our result in terms of kinetic potentials and prove (for the case $\ell = 0$) the following:\medskip
% -----------------
\nll{\bf Theorem~1}~~
% -----------------
{\th If $E$ is the bottom of the spectrum of the Hamiltonian $H = -\Delta + V,$ 
and the potential $V$ is the sum $V(r) = h^{(1)}(r) + h^{(2)}(r),$ then it follows that the sum of the component kinetic potentials yields a lower bound to $\wbar{V},$ that is to say}
$$\wbar{V}(s) \geq \wbar{h}^{(1)}(s) + \wbar{h}^{(2)}(s).\eqno{(3.1)}$$
\nl We shall now prove this theorem, which is in effect an optimized Weyl lower bound\sref{\hallprd--\weyl}. 
 From the definition (1.1) of kinetic potentials we have
$$\wbar{V}(s) = \inf_{{{\scriptstyle \psi \in {\cal D}(H)} \atop {\scriptstyle (\psi,\psi) = 1}} \atop {\scriptstyle (\psi, K\psi) = s}} (\psi, V\psi) 
= \inf_{{{\scriptstyle \psi \in {\cal D}(H)} \atop {\scriptstyle (\psi,\psi) = 1}} \atop {\scriptstyle (\psi, K\psi) = s}} \left(\psi, \left(h^{(1)} + h^{(2)}\right)\psi\right).$$
\nl But the latter minimum mean-value is clearly bounded below by the sum of the {\it separate}
 minima.  Thus we have  
$$\wbar{V}(s) \geq \inf_{{{\scriptstyle \psi \in {\cal D}(H)} \atop {\scriptstyle (\psi,\psi) = 1}} \atop {\scriptstyle (\psi, -\Delta\psi) = s}} \left(\psi, h^{(1)}\psi\right) 
+  \inf_{{{\scriptstyle \psi \in {\cal D}(H)} \atop {\scriptstyle (\psi,\psi) = 1}} \atop {\scriptstyle (\psi, -\Delta\psi) = s}} \left(\psi, h^{(2)}\psi\right) =
 \wbar{h}^{(1)}(s) + \wbar{h}^{(2)}(s),$$
\nl which inequality establishes the theorem.\qed

Another approach, which would eventually yield an alternative 
proof of the theorem, exhibits the relationship between Theorem~1 and the classical Weyl lower bound\sref{\weyl--\wein} for the eigenvalues of the sum of two operators.  Let us suppose that $\Psi$ is the exact normalized lowest eigenfunction of $H = -\Delta + V,$ so that $H\Psi = E\Psi.$  If the positive real parameter $w$ satisfies $0 < w < 1,$ then $E = (\Psi, (-\Delta + V)\Psi)$ may
be written as follows: 
$$\eqalign{E & =\  w\left(\Psi,\left(-\Delta + {1\over w}\ h^{(1)}(r)\right)\Psi\right)
+ (1-w)\left(\Psi, \left(-\Delta + {1\over{1-w}}\ h^{(2)}(r)\right)\Psi\right)\cr
& \geq\  w\inf_{{{\scriptstyle \psi \in {\cal D}(H)} \atop {\scriptstyle (\psi,\psi) = 1}}}
\left(\psi,\left(-\Delta + {1\over w}\ h^{(1)}(r)\right)\psi\right)\cr
&+\  (1-w)\inf_{{{\scriptstyle \psi \in {\cal D}(H)} \atop {\scriptstyle (\psi,\psi) = 1}}}
\left(\psi,\left(-\Delta + {1\over{1-w}}\ h^{(2)}(r)\right)\psi\right).}$$
\nl That is to say, in terms of component kinetic potentials, we arrive at Weyl's inequality
for the lowest eigenvalue $E$ of the operator sum $H,$ where
$$H = -w\Delta + h^{(1)}\quad +\quad -(1-w)\Delta + h^{(2)},$$ 
and we conclude
$$E \geq w\min_{s >0}\left\{s + {1\over w}\ \wbar{h}^{(1)}(s)\right\} + 
(1-w)\min_{s >0}\left\{s + {1\over{1-w}}\ \wbar{h}^{(2)}(s)\right\}.$$
\nl Since $w$ is an essentially free parameter in the last expression, we may optimize
the Weyl lower bound with respect
to the choice of $w$: this forces the individual values of $s$ at the minima,
 $\{s_1(w), s_2(w)\},$ to be related.  More specifically we find from the
individual minimizations over $s,$
$$E \geq {\cal E}(w) = ws_1(w) + (1-w)s_2(w) + \wbar{h}^{(1)}(s_1(w))+ \wbar{h}^{(2)}(s_2(w)),$$
\nl where
$$w = -{{\partial\wbar{h}}\over{\partial s}}^{(1)}(s_1(w)),\quad{\rm and}\quad 1-w = -{{\partial\wbar{h}}\over{\partial s}}^{(2)}(s_2(w)).$$
\nl The critical condition ${\cal E}^{\prime}(w) = 0$ for the subsequent maximization
 over $w$ then yields $s_1(w) = s_2(w).$ Thus the best lower energy bound is 
given by
$$E \geq \min_{s >0}\left\{s + \wbar{h}^{(1)}(s)+ \wbar{h}^{(2)}(s)\right\}.$$
\nl The kinetic-potential inequality of Theorem~1 leads, of course, to the same energy lower bound: the optimization just performed above is therefore seen to be automatically `built in'
by the kinetic-potential formalism.  

It follows immediately from the above kinetic-potential comparison theorem 
and coupling-parameter absorption that a lower bound to the lowest energy $E$ of
the Hamiltonian $H = -\Delta + \sum_{i}c_{i}h^{(i)}(r),$ $\{c_i >0\},$ is provided by the formula
$$E\  \geq\  \min_{s >0}\left\{s + \sum_{i}c_{i}\wbar{h}^{(i)}(s)\right\}.\eqno{(3.2)}$$
\nl Similarly we can extend this result to `continuous sums' such as 
$V(r) = \int_{t_1}^{t_2} c(t)h^{(t)}(r)dt.$

Meanwhile, since the proof is identical, the bound is valid for the bottom of each angular-momentum subspace. Thus, more generally, the fundamental inequality becomes
$$\wbar{V}_{1\ell}(s) \geq \wbar{h}_{1\ell}^{(1)}(s) + \wbar{h}_{1\ell}^{(2)}(s),\  \ell = 0,1,2,\dots.\eqno{(3.3)}$$
%---------------------------------------------------------------
\title{4.~~Generalized Comparison Theorems}
%---------------------------------------------------------------
The proof of our generalized comparison theorem (Theorem~4) depends on monotone behaviour of the wave function induced by the assumed monotonicity of the potential.  We are able to establish this monotonicity for the lowest eigenfunction in arbitrary many spatial dimensions $N \geq 1.$ We shall then be able to apply our eigenvlaue results to the case $\ell >0$ and $n = 1$ because of Theorem~2 which claims that $E_{n\ell}^{N} = E_{n 0}^{N+2\ell};$ this general result is then employed in the special case $n = 1.$

\noindent {\bf Theorem 2}

{\th \noindent Suppose that $H = -\Delta +V(r),$ where $V(r)$ is a central potential in $N\geq 2$ dimensions, has a discrete eigenvlaue $E_{n\ell}^{N}$ with $n$ radial nodes in the angular-momentum subspace labelled by $\ell,$ then  $E_{n\ell}^{N} = E_{n 0}^{N+2\ell}.$}
\nll{\bf Proof:} We suppose that $\psi$ is the eigenfunction corresponding to  $E_{n\ell}^{N}.$
We express $-\Delta $ in spherical coordinates\sref{\som-\schl} and write the radial 
eigenequation explicitly as
$$-\psi''(r)-{(N-1)\over r}\psi'(r)+{\ell(\ell+N-2)\over r^2} \psi(r)+V(r)\psi(r)=E^N_{n\ell}\psi(r). $$

\noindent If we now define the reduced radial function $u(r)\in L^2(R^{+})$ by $\psi(r)=u(r)r^{-(N-1)/2},$  $r > 0,$ and $u(0) = 0,$ we obtain

$$-u''(r)+\left[{{ {{(N-1)}\over{2}}{(N-3)\over{2}}+\ell(\ell+N-2)}\over {r^2}}+V(r)\right]u(r)=E^N_{n\ell}u(r).\eqno{(4.1)} $$
\noindent If we consider the spherically-symmetric potential $V(r)$ in $M$ dimensions such that $(M-1)(M-3)/4=\ell(\ell+N-2)+(N-1)(N-3)/4 $, we find that $M=2\ell+N.$ The eigenequation (4.1) then may be written equivalently
$$-u''(r)+\left[{{{(M-1)\over 2}{(M-3)\over 2}}\over r^2}+V(r)\right]u(r)={{E}}^{N}_{n\ell}u(r).\eqno{(4.2)}$$
\nl It therefore follows immediately that $E^N_{n\ell}=E^{M}_{n0}=E^{2\ell+N}_{n0}$.\qed 

%---------------------------------------------------------
For the purpose of our comparison theory we may now consider the special case $n=1,~\ell=0$ in arbitrary $N \geq 1$ spatial dimensions: the energy results which we derive will then be applicable to the family of equivalent problems in $N'$ spatial dimension with $n=1,$ $\ell > 0,$ and $N = N'+2\ell.$  In order to prove an appropriate extension of the comparison theorem in $N$ dimensions, we shall first need to establish an elementary monotonicity property for the ground-state $\psi.$ We prove the following:

\noindent {\bf lemma }

\noindent {\th Suppose $\psi=\psi(r),$ $r=\|\mb{r}\|,$ $\mb{r}\in R^{N},$ satisfies Schr\"odinger's equation:
$$H\psi(r)=(-\Delta+V(r))\psi(r)=E\psi(r),\eqno{(4.3)}$$
\nl where $V(r)$ is a central potential which is monotone increasing for $r > 0.$  Suppose that $E$ is a discrete eigenvalue at the bottom of the spectrum of the operator $H = -\Delta + V,$ defined on some suitable domain ${\cal D}(H)$ in $L^{2}(R^N).$ Suppose that $\psi(r)$ has no nodes, so that, without loss of generality, we can assume $\psi(r) > 0,\quad r > 0.$ Then $\psi'(r) \leq 0,\quad r > 0.$}

\noindent{\bf Proof:} The proof for the case $N=1$ is given in Ref.\sref{\hallt}, Eq.(2.2).  Henceforth we shall now assume $N\geq 2.$ If we express $-\Delta$ in spherical coordinates in $N$ spatial dimensions, then we have 
$$-\Delta \psi +V \psi =E\psi$$
$$-{1\over {t^{N-1}}}{\partial\over {\partial t}}(t^{N-1}{\partial \over {\partial t} }) \psi(t)+V(t)\psi(t)=E\psi(t)   $$
\noindent We now multiply by $t^{N-1}$ both sides and integrate with respect to $t$, to obtain 
$$\psi'(r)=(1/r^{N-1}) \int_0^r [V(t)-E]\psi(t)t^{N-1} dt. $$
\noindent Since $V$ is monotone increasing, it follows that there is one point $\hat{r} > 0$ satisfying $V(\hat{r}) = E.$ 
\noindent First, we prove that $s(r)=\int_{\hat{r}}^r[V(t)-E]\psi(t)t^{N-1}dt$ is monotone increasing and bounded. For $t>\hat{r},$  $[V(t)-E]\psi(t)t^{N-1}>0,$ because $V(t)>E$ and hence $s(r)$ is increasing as $r\rightarrow \infty$. If there exists $r_1<\infty$ such that $s(r_1)=-\int_0^{\hat{r}}[V(t)-E]\psi(t)t^{N-1}dt$, then $\int_0^{r}[V(t)-E]\psi(t)t^{N-1}dt>0,~r>r_1$, and  $\psi'(r)>0,~r>r_1$; this contradicts the fact that the wave function $\psi(r)$ is positive and belongs to $L^{2}(R^N)$. This means that $\int_0^r[V(t)-E]\psi(t)t^{n-1}dt\le 0,~\forall r>0$. Consequently $\psi'(r)\le 0,~\forall r>0.$\qed 

We now consider two potentials $V_1(r)$ and $V_2(r)$ both of the type described above. We have two Sch\"odinger equations for the respective ground-states $\psi_1$ and $\psi_2$ and the corresponding discrete eigenvalues $E_1$ and $E_2$ at the bottoms of the spectra. Thus we have the following pair of eigenequations
$$(-\Delta +V_1(r))\psi_1(r)=E_1\psi_1(r)\eqno{(4.4)} $$
$$(-\Delta +V_2(r))\psi_2(r)=E_1\psi_2(r)\eqno{(4.5)} $$
 \noindent  The radial wave functions in the present paper satisfy the normalization condition  
$\int_0^\infty \psi_i^2(r)r^{N-1}dr<\infty,$ $i = 1,2.$  With this notation, and $N=3,$ Theorem~3 of Ref.\sref{\hallt} becomes

%--------------------------
\nl{\bf Theorem 3}
% -------------------------
{\th $$k(r)=\int_0^r(V_1(t)-V_2(t))\psi_i(t)t^2dt\le 0,~\forall r>0,~i=1~{\rm or}~2 \Rightarrow E_1\le E_2.\eqno{(4.6)}$$ }

\noindent We shall now generalize this theorem to general dimension $N\ge 1$.
 We first establish a fundamental comparison formula (Eq.(4.7)) below.
 
\noindent By multiplying $(4.4)$ by $\psi_2$ and $(4.5)$ by $\psi_1,$ and subtracting, we find  

$$\psi_1\Delta\psi_2-\psi_2\Delta\psi_1+[V_1-V_2]\psi_1\psi_2=[E_1-E_2]\psi_1\psi2 $$

\noindent Integrating over $R^N$ and using the following identity,

$$\nabla.(\psi_1\nabla\psi_2)=\nabla\psi_1.\nabla\psi_2+\psi_1\nabla^2\psi_2$$  

\noindent we find that,

$$\int_{R^N}\nabla.[\psi_1\nabla\psi_2-\psi_2\nabla\psi_1] d^Nr+\int_{R^N}[V_1-V_2]\psi_1\psi_2d^Nr=[E_1-E_2]\int_{R^N}\psi_1\psi_2d^Nr $$

\noindent Now by Gauss's theorem\sref{\corant} we find that the first term becomes a surface integral which vanishes because $\psi_i\in L^2(R^N).$ In the remaining integrals the angular factors yield $2\pi^{N/2}/\Gamma(N/2).$ 
 Hence we find 
 
$${2\sqrt{\pi^N}\over \Gamma(N/2)}\int_0^{\infty}[V_1(r)-V_2(r)]\psi_1(r)\psi_2(r) r^{N-1}dr={2\sqrt{\pi^N}\over \Gamma(N/2)}[E_1-E_2]\int_0^{\infty}\psi_1(r)\psi_2(r)r^{N-1}dr, $$

\noindent which implies,

$$ s=\int_0^\infty[V_1-V_2]\psi_1\psi_2r^{N-1}dr=[E_1-E_2]\int_0^\infty\psi_1\psi_2r^{N-1}dr\eqno{(4.7)}$$

\noindent Now we may state our generalization of Theorem 3 to $N$ dimensions:

\noindent{\bf Theorem 4}
$$k(r)=\int_0^r(V_1(t)-V_2(t))\psi_i(t)t^{N-1}dt\le 0,~\forall r>0,~i=1~{\rm or}~2 \Rightarrow E_1\le E_2.\eqno{(4.8)}$$ 

\noindent{\bf Proof:} For definiteness we assume that $i=1;$ the proof is just the same with the other choice. We study the integral $s$ on the left side of $(4.7)$. Integrating by parts we find that

$$s=[k(r)\psi_2(r)]_0^\infty-\int_0^\infty k(r)\psi_2'(r) r^{N-1}dr \eqno{(4.9)} $$
\noindent Since $k(0)=\psi_2(\infty)=0,$ the first term vanishes, and $s$ is therefore equal to the negative of the integral of the right side of $(4.9)$. But the integrand of this integral is positive because $k(r)\le 0,$ by hypothesis,  and we know that $\psi_2'(r)\le 0$ by the above lemma. This proves that $s\le 0.$ Consequently, by $(4.7),$ we obtain $E_1\le E_2$.
 \qed
%----------------------------------------------------------

%----------------------------------------------------------
It may be difficult to apply Theorem 4 in practice. Thus it would be helpful to establish some simpler sufficient conditions, depending on the number and nature of the crossings over of the two comparison potentials. We treat three useful cases: Theorem 5, one potential crossing, with use of the wave function; Theorem 6, two crossings and the use of the wave function; Theorem 7, two crossings and no wave function used. In these Theorems we shall assume that the integrals $\int_0^\infty(V_1(r)-V_2(r))\psi_i(r)r^{N-1}dr,$ $i = 1,2,$ exists for the given problem, even though we use at most one wave function factor.

%-------------------------
\noindent{\bf Theorem 5.}~~
%-------------------------
{\th If the potentials $V_1(r)$ and $V_2(r)$ cross exactly once for $r>0$ at $r=r_1$, with,

\noindent (i) $V_1(r)<V_2(r)$ $(0<r<r_1)$ and 

\noindent (ii) $\int_0^\infty [V_1(t)-V_2(t)]\psi_i(t)t^{N-1}dt \leq 0,~i=1~{\rm or}~2, $ 

\noindent then
$$k(r)=\int_0^r[V_1(t)-V_2(t)]t^{N-1}\psi_i(r)dt\le 0 ,~\forall~ r>0,\ i = 1\ {\rm or}\ 2,\eqno{(4.10)}$$ 

\noindent from which $E_1\le E_2$ follows, by Theorem 4.
}
\nll{\bf Remark:} The best bound is obtained with the equality in hypothesis (ii).
\medskip
\nl{\bf Proof of Theorem 5:} We choose $i=1:$ the proof is identical for $i=1$ or $2.$ First, we show that $s(r)=\int_{r_1}^r[V_1(t)-V_2(t)]\psi_1(t)t^{N-1}dt$ is monotone increasing. For $t>r_1,$  $s'(r)= [V_1(r)-V_2(r)]\psi_1(r) r^{N-1}>0,$ because $V_1(r)>V_2(r);$ hence $s(r)$ is increasing on $(r_1,\infty)$. Moreover, (ii) implies that the maximum value of $s(r)$ is reached at $r=\infty;$ i.e $s(r)\le s(\infty)$ we have therefore

$$\int_0^\infty[V_1(t)-V_2(t)]\psi_1(t)t^{N-1}dt=$$ 
$$\int_0^{r_1} [V_1(t)-V_2(t)]\psi_1(t)t^{N-1}dt +\int_{r_1}^\infty [V_1(t)-V_2(t)]\psi_1(t)t^{N-1}dt \leq 0$$

\noindent and therefore  

$$ \lim_{r\rightarrow \infty}s(r) \leq -\int_0^{r_1} [V_1(t)-V_2(t)]\psi_1(t)t^{N-1}dt. $$ 

\noindent Now, we have the following two cases to consider 

\noindent Case 1: for $r<r_1$, $ k(r)=\int_0^r [V_1(t)-V_2(t)]\psi_1(t)t^{N-1}dt \leq 0,$ since $V_1(t)<V_2(t)$ for $0<t<r.$ 

\noindent Case 2: if $r>r_1$, then $$ k(r)=\int_0^{r_1}[V_1(t)-V_2(t)]\psi_1(t)t^{N-1}dt +\int_{r_1}^r[V_1(t)-V_2(t)]\psi_1(t)t^{N-1}dt=s(r)-s
(\infty)<0$$  
 \noindent Therefore, $k(r)\le 0,\forall r>0.$\qed\medskip

%---------------------------
\noindent{\bf Theorem 6.}~~
%---------------------------
 {\th If the potentials $V_1(r)$ and $V_2(r)$  cross  twice for $r>0$ at $r=r_1$,$r=r_2$ $(r_1<r_2)$ with,

\noindent (i) $V_1(r)<V_2(r)$ for $0<r<r_1$ and 

\noindent (ii) $\int_0^{r_2} (V_1(t)-V_2(t))\psi_i(t)t^{N-1}dt \leq 0,\ i = 1\ {\rm or}\ 2, $  

\noindent then,
$$k(r)=\int_0^r(V_1(t)-V_2(t))\psi_i(t)t^{N-1}dt\le 0,~\forall r>0,\ i = 1\ {\rm or}\ 2,\eqno{(4.11)}$$ 

\noindent from which $E_1\le E_2$ follows, by Theorem 4.
}
\nll{\bf Remark:} The best bound is obtained with the equality in hypothesis (ii).
\medskip
\nl{\bf Proof of Theorem 6:} $k'(r)=(V_1(r)-V_2(r))r^{N-1}\psi_1(r)$, now $k(0)=0,$ $k'(r)<0,~0<r<r_1,$ implies $k(r)<0,$ $0<r<r_1.$ Next, $k(r_2)=0,$ $k'(r)>0,~r_1<r<r_2,$ implies $k(r)<0,~r_1<r<r_2.$ Lastly, $k(r_2)=0,~k'(r)<0,~r>r_2,$ implies $k(r) \leq 0,~r>r_2.$\qed
\medskip

%------------------------
\noindent{\bf Theorem 7.}~~
%------------------------
{\th If the potentials $V_1(r)$ and $V_2(r)$ cross  twice for $r>0$ at $r=r_1,~r_2$ $(r_1<r_2)$ with

\noindent(i) $V_1(r)<V_2(r)$ for $0<r<r_1$ and 

\noindent(ii) $\int_0^{r_2} (V_1(t)-V_2(t))t^{N-1}dt \leq 0 $
 
\noindent then,  
$$k(r)=\int_0^r[V_1(t)-V_2(t)]\psi_i(t)t^{N-1}dt\le 0 ,~\forall r>0~,\ 
 i = 1\ {\rm or}\ 2, \eqno{(4.12)}$$ 
 \noindent from which $E_1<E_2$ follows, by Theorem 4.}
\nll{\bf Remark:} The best bound is obtained with the equality in hypothesis (ii).
\medskip
\nl{\bf Proof of Theorem 7:} We choose $i = 1:$ the proof is identical if $i = 2.$  Define $h(r)=\int_0^r(V_1(t)-V_2(t))t^{N-1} dt$ the proof of Theorem 3 shows that $h(r)\le 0,~0\le r\le r_2.$ But
$$k(r)=\int_0^r(V_1(t)-V_2(t))\psi_1(t)t^{N-1}dt $$
$$=\left[h(t)\psi_1(t)\right]_0^r-\int_0^rh(t)\psi_1'(t)dt $$
$$=h(r)\psi_1(r)-\int_0^rh(t)\psi_1'(t)dt<0, $$
\noindent meanwhile, $k'(r)<0,~r>r_2.$ Therefore, $k(r) \leq 0,~\forall r>0.$ \qed
\medskip

%---------------------------------------------------------------
\title{5.~~Application to the Coulomb-plus-linear potential}
%---------------------------------------------------------------
As an example, we employ the comparison theorems to improve the bounds obtained in Section~2
for the eigenvalues corresponding to the Coulomb-plus-linear potential $V(r)=-a/r + b r$, where $a$ and $b$ are positive coupling parameters.
For the upper bound we use as a comparison potential the shifted linear potential $h(r)=-\alpha+\beta r,$ where $\alpha~{\rm and }~\beta >0$.  We allow the potentials $V(r)$ and $h(r)$ to cross over exactly twice, as illustrated in Figure~2. Let $A$ and $B$ represent the absolute values of the areas (or of the $\psi$-weighted areas) between the potentials. We vary $\alpha$ and  $\beta$ so that $A=B$, and thereafter Theorems 5 and 6 imply $E^V\le E^h.$ For simplicity of derivation of the upper-bound formula, we will use Theorem 7 (with no use of the wave function $\psi$). Thus we have two equations to solve in this case,
$$-{a\over r}+br=-\alpha+\beta r, $$
$$\int_0^{r}[-{a\over t}+bt+\alpha-\beta t]t^{N-1}dt=0, $$
\noindent where $r=r_2$ is the second crossing point. These reduce to the quadratic equations
$$(\beta-b)r^2-\alpha r+a=0, $$
$$N(N-1)(b-\beta)r^2+\alpha(N-1)(N+1) r-aN(N+1)=0, $$
\nl with simultaneous solution $r={2aN\over {\alpha (N-1)}}$. Now the best upper bound is
obtained after minimizing with respect to r, giving
$$E^U=\min_{r>0}\left\{ -\left({2aN\over{(N-1) r}}\right)+\left({(N+1)a\over {(N-1)r^2}}+b\right)^{2\over 3}{\cal{E}}^N(1)\right\}.\eqno{(5.1)}$$
\nl At the expense of further complication, the use of $\psi_1(r)$ (the Airy function) would lower this upper bound.

  Similarly, to improve our lower bound, we allow the Coulomb-plus-linear potential to intersect twice with the Hydrogenic potential $h(r)=-{\alpha\over r}+\beta$, with the exact wave function solution $\psi=e^{-\alpha r/(N-1)}$ and the exact energy $E^h=\beta-\alpha^2/(N-1)^2,$ where $\alpha$ and $\beta$ are positive parameters. Again, let $A$ and $B$ represent the absolute values of the areas (or of the $\psi$-weighted areas) between the potentials. We vary $\alpha$ and  $\beta$ so that $A=B$, as illustrated in Figure~3, and thereafter Theorems~5 and 6 imply $E^h\le E^V.$ Subsequently, we obtain the lower bound (without $\psi$) by solving the following three equations:
$${-a\over t}+b t=-{\alpha\over t}+\beta $$
$$\int_0^t[{-a\over r}+b r+{\alpha\over r}-\beta]r^{N-1}dr=0 $$
$$E^L=\min_{t>0}\{\beta-{\left(\alpha /(N-1)\right)}^2 \}.\eqno{(5.2)}$$
\nl For the case $a=1$ and $b=1$, we compare in Figure~4 the upper and lower bounds obtained by means of the classical envelope method and by the comparison theorems introduced in Section~3.
Generalizations to cases where there are a large number of potential crossings are discussed in Ref.\sref{\hallns}.
% ------------------------------------------------------
  \title{6.~~Conclusion}
% ------------------------------------------------------

Our proof of the lower-bound for the bottom of the spectrum of the 
operator $H = -\Delta + V^{(1)}(r)+V^{(2)}(r),$ based on kinetic potentials, 
is more compact and direct than the original proof, and is valid in $N$ dimensions;
 the principal steps of the earlier
proof are repeated because they show that the final result is equivalent to an
optimization of the classical theorem of Weyl. The generalized comparison theorem
is proved in the present paper for all dimensions $N,$ whereas,
 in its original form, it required two
distinct theorems, for $N=1,$ and $N = 3.$  Moreover, we are now able to apply the results to the bottom of each angular-momentum subspace since we have proved that this energy is identical 
to the lowest eigenvalue of a higher-dimensional problem, in $R^{N+2\ell}.$   Meanwhile, in order to be practical, weaker sufficient conditions were sought which would guarantee in a simple way that the comparison potentials cross over so as to imply definite spectral ordering.  These results greatly clarify the application of the generalized comparison theorem to specific problems. 

The Coulomb-plus-linear problem provides a convenient example on which to test the effectiveness of the energy bounds.  At the same time it offers an opportunity to sharpen an earlier energy-bound formula for this problem, and to extend its validity to all $N\geq2$ dimensions.  For most of the parameter space of the problem, the energy bounds provided by this formula for the bottom of each angular-momentum subspace $(n = 1)$ are accurate to a few percent and, as we have shown, they become sharper with increasing $N$ or $\ell.$  If the sum approximation is capriciously applied also to the higher discrete eigenvalues $n > 1$, the resulting {\it ad hoc} approximation formula continues to give
very accurate estimates, which, however, are no longer bounds.  What additional conditions might
guarantee bounds from such a formula is an interesting open question.

% ------------------------------------------------------   
   \title{Acknowledgments}
% ------------------------------------------------------
Partial financial support of this work under Grant No.GP3438 from the Natural Sciences and Engineering Research Council of Canada is gratefully acknowledged.\bigskip   
% ------------------------------------------------------ 
\np
\references{1}     
\np
% ------------------------------------------------------
%  Table
% ------------------------------------------------------
\noindent {\bf Table 1}~~The `input' \hi{P}{values} $P^N_{n 0}(1)$  
used in the general formula (1.4), for $N=2,3,\dots,12$. The same data applies
to $\ell > 0$ since, by Theorem~2, we have $P_{n\ell}^{N} = P_{n 0}^{N+2\ell}.$ 

\baselineskip=16 true pt % adjusted for table
\def\vr{\vrule height 12 true pt depth 6 true pt}
\def\vra{\vr\hfill} \def\vrb{\hfill &\vra} \def\vrc{\hfill & \vr\cr\hrule}
\def\vrq{\vr\quad} 

$$\vbox{\offinterlineskip
 \hrule
\settabs
\+ \vrq \kern 0.9true in &\vrq \kern 0.9true in &\vrq \kern 0.9true in &\vrq \kern 0.9true in &\vrq \kern 0.9true in&\vrq \kern 0.9true in&\vrq \kern 0.9true in&\vrq \kern 0.9true in&\vr\cr\hrule
\+ \vra $N$ \vrb  $n=1$\vrb $n=2$\vrb $n=3 $\vrb $n=4 $  \vrc
%\+ \vra 1\vrb  1.37608\vrb 1.8735\vrb 2.3719\vrb 2.8709\vrc
\+ \vra 2 \vrb 0.9348\vrb    2.8063\vrb    4.6249\vrb    6.4416\vrc
\+ \vra 3 \vrb 1.3761 \vrb   3.1813 \vrb   4.9926 \vrb   6.8051\vrc
\+ \vra  4 \vrb  1.8735\vrb    3.6657\vrb    5.4700\vrb    7.2783\vrc
\+ \vra  5\vrb  2.3719\vrb    4.1550\vrb    5.9530\vrb    7.7570\vrc
\+ \vra 6\vrb   2.8709\vrb    4.6472\vrb    6.4398\vrb    8.2396\vrc
\+ \vra 7\vrb   3.3702\vrb    5.1413\vrb    6.9291\vrb    8.7251\vrc
\+ \vra 8\vrb   3.8696\vrb    5.6367\vrb    7.4204\vrb    9.2129\vrc
\+ \vra 9\vrb   4.3692\vrb    6.1330\vrb    7.9130\vrb    9.7024\vrc
\+ \vra 10\vrb   4.8689\vrb    6.6299\vrb    8.4068\vrb   10.1932\vrc
\+ \vra 11 \vrb 5.3686\vrb    7.1274\vrb    8.9053\vrb   10.7453\vrc
\+ \vra 12\vrb   5.8684\vrb    7.6253 \vrb   9.4045\vrb   11.2744\vrc
}$$
%% -------------------------------------------------------------------

\np

\hbox{\vbox{\psfig{figure=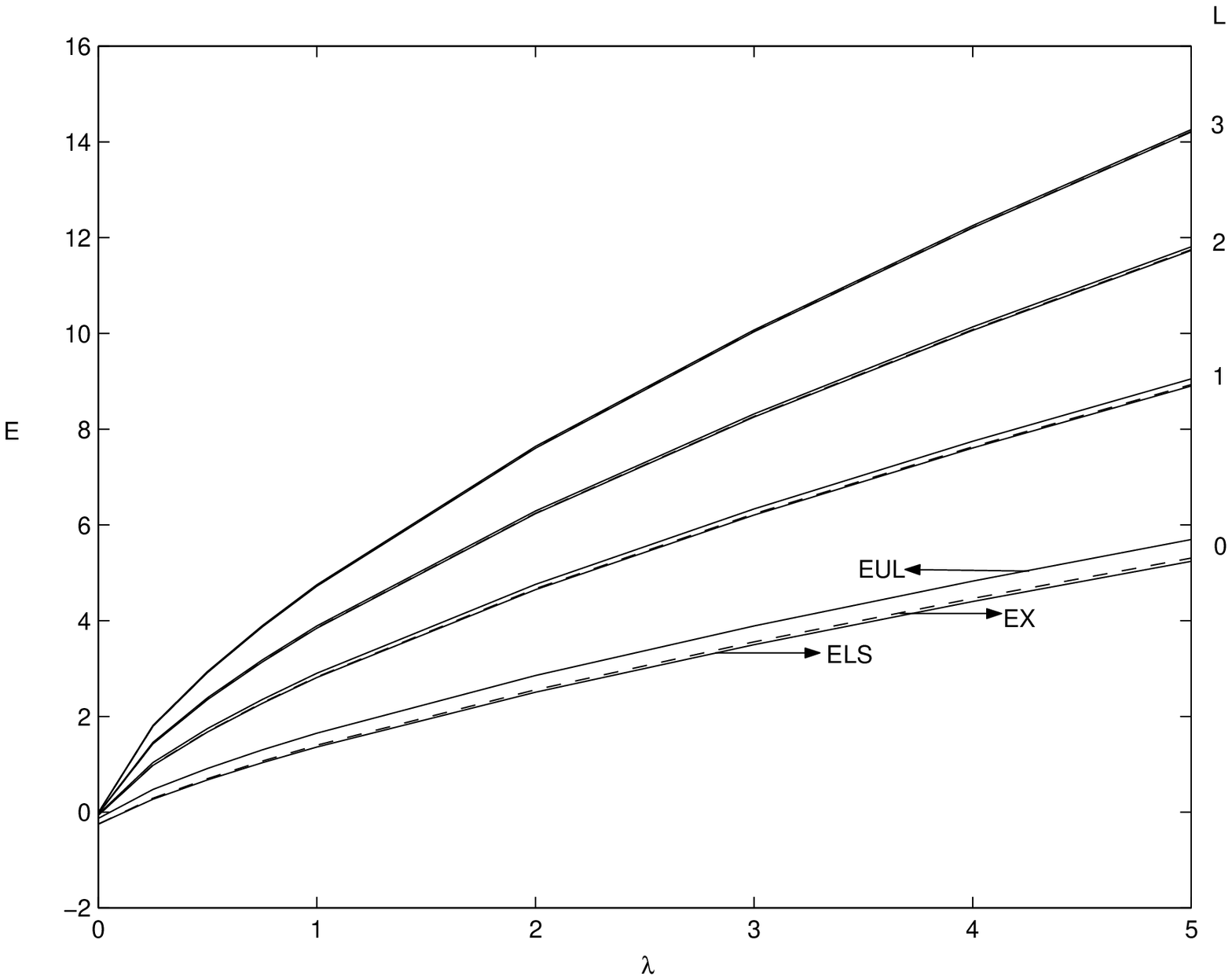,height=4in,width=5in,silent=}}}

\title{Figure 1.}

\nl The eigenvalues ${E}(\lambda)$ of the Hamiltonian $H=-\Delta-1/r+\lambda r$ for $N = 3,$ $n=1,$ and $\ell= L =0,1,2,3$. The continuous curves show the upper bound EUL given by the envelope formula (2.9) 
with $\nu = \mu = P_{1\ell}^{3}(1),$ and the lower bound ELS by the sum approximation 
given by the same formula but with $\nu = P_{1\ell}^{3}(1)$ and $\mu = P_{1\ell}^{3}(-1).$  The dashed curve EX represents accurate numerical data.

%\medskip
% -----------------------------------------------------------------------
\np
\hbox{\vbox{\psfig{figure=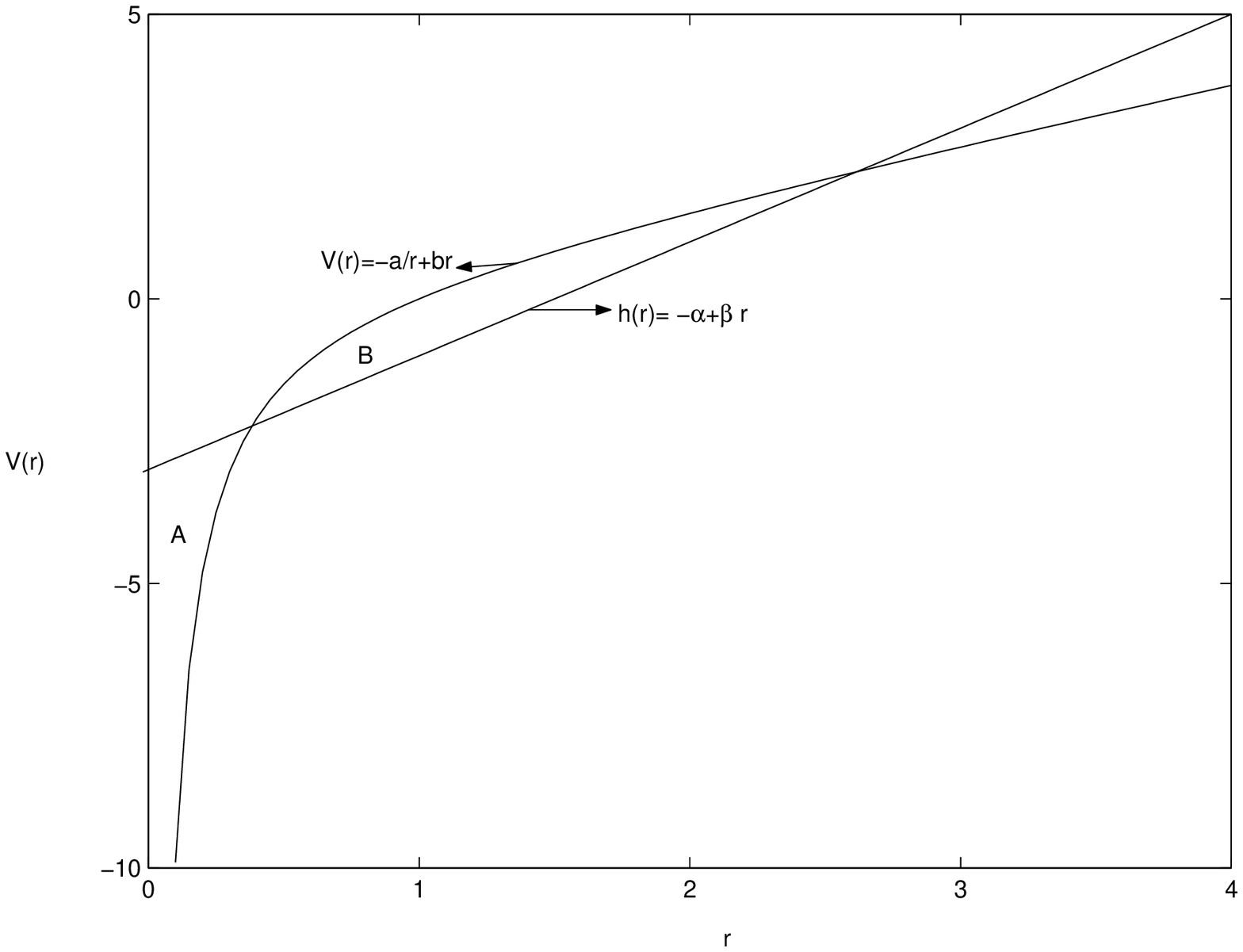,height=4in,width=5in,silent=}}}
\title{Figure 2.}
\nl The linear potential $h(r)=\alpha r+\beta$ used to estimate an upper bound for the eigenvalues of the Coulomb-plus-linear potential $V(r)=-a/r+br$. $A$ and $B$ are the absolute values of the inter-potential areas (or $\psi$-weighted areas). We vary $\alpha$ and $\beta$ so that $A=B$, and thereafter Theorems 5 and 6 imply $E^V\le E^h.$    
% -----------------------------------------------------------------------
\np
\hbox{\vbox{\psfig{figure=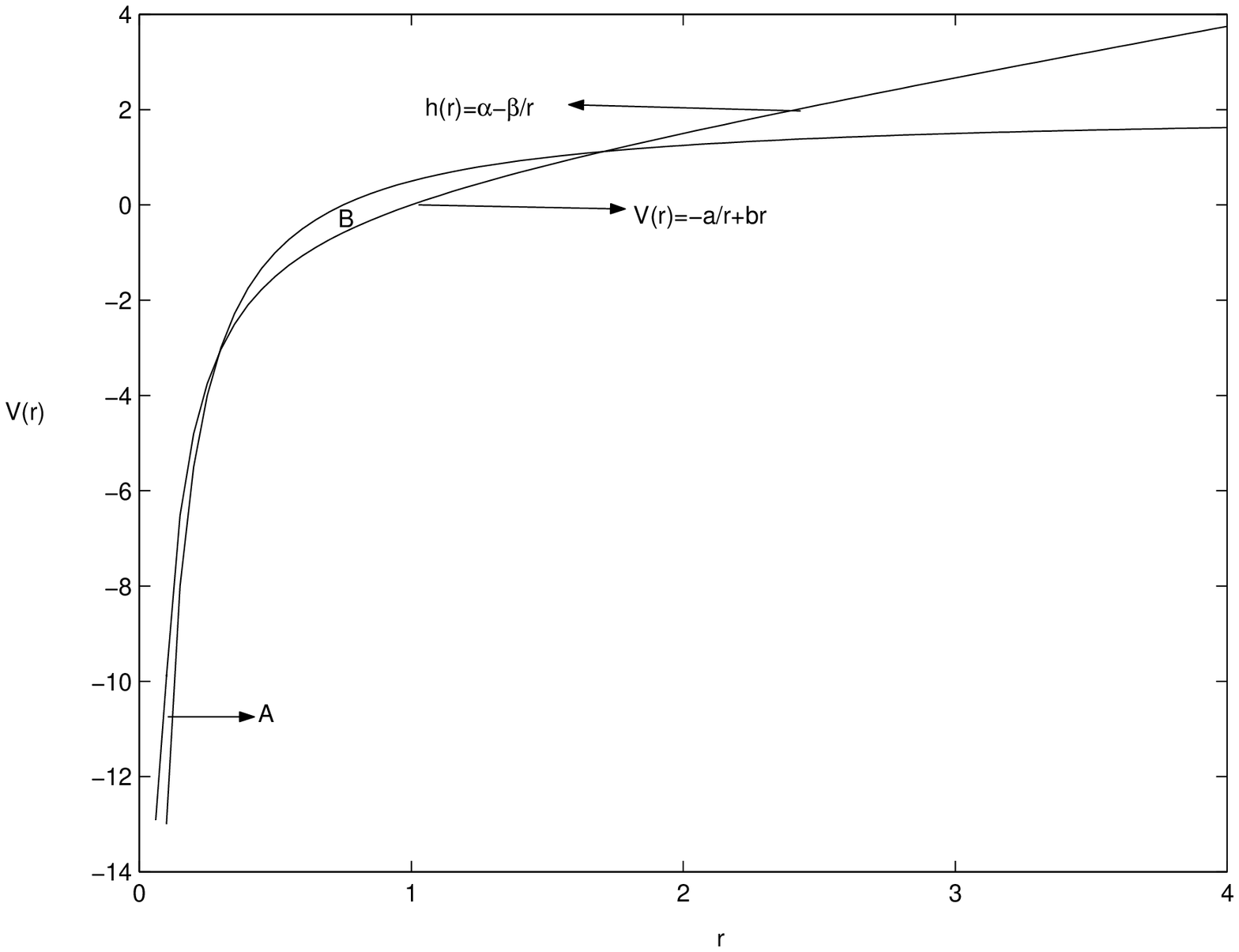,height=4in,width=5in,silent=}}}
\title{Figure 3.}
\nl The hydrogenic potential $h(r)=-\alpha/r+\beta$ used to estimate a lower bound for the eigenvalues of the Coulomb-plus-linear potential $V(r)=-a/r+br$. $A$ and $B$ are the absolute values of the inter-potential areas (or $\psi$-weighted areas). We vary $\alpha$ and $\beta$ so that $A=B$, and thereafter Theorems 5 and 6 imply $E^h\le E^V.$  
%\medskip
% -------------------------------------------------------------------------
\np
\hbox{\vbox{\psfig{figure=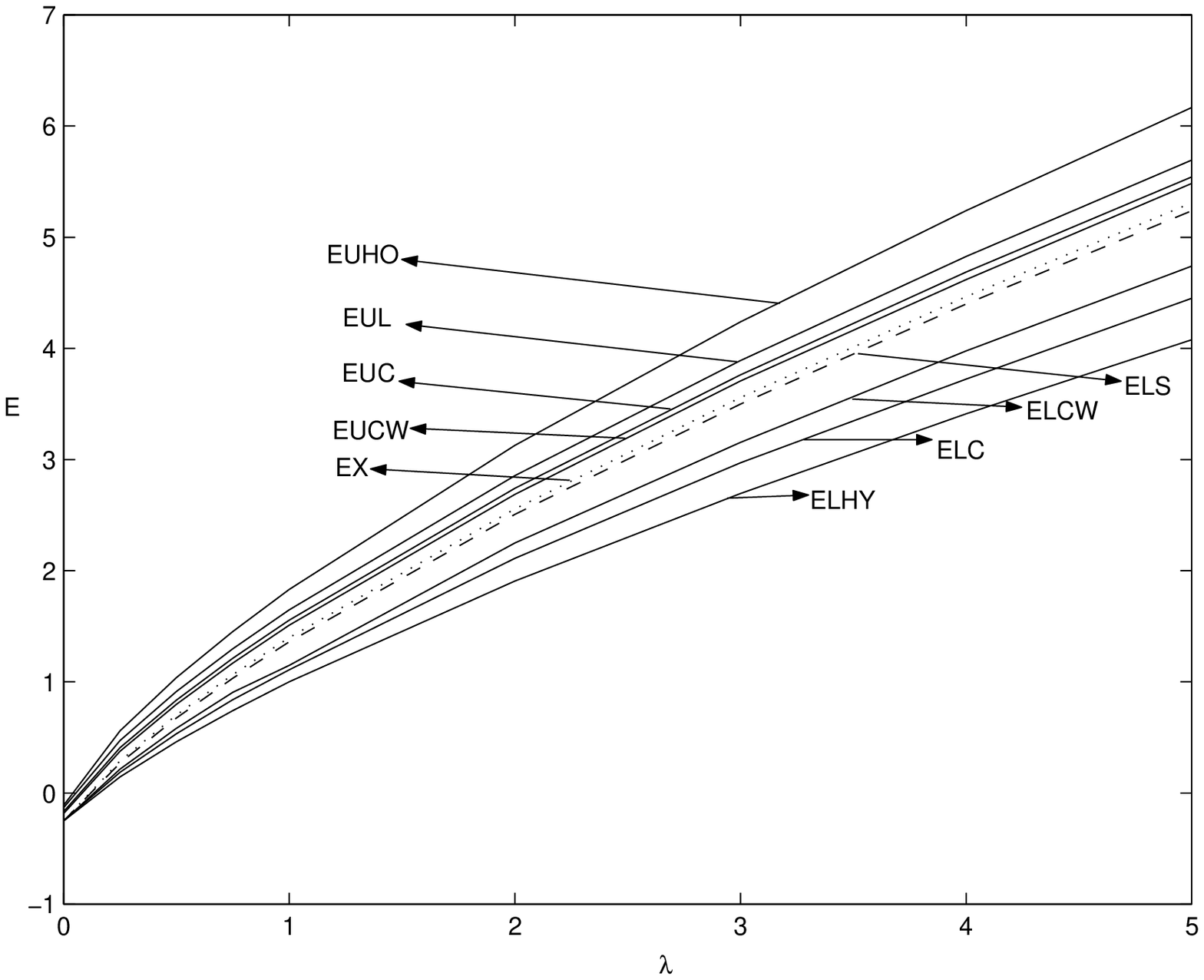,height=4in,width=5in,silent=}}}
\title{Figure 4.}
\nl We compare the bounds for ${E}(\lambda)$, where ${E}(\lambda)$ is the ground-state eigenvalue $(n = 1,~ \ell = 0)$ of the Hamiltonian $H=-\Delta-1/r+\lambda r$. The upper bounds (full-line) are by harmonic-oscillator tangents EUHO,  linear tangents EUL, linear chords EUC, and linear chords with the
wave function EUCW. The lower bounds (lower full-lines) are by hydrogenic tangents ELHY, Hydrogenic chords ELC, and 
Hydrogenic chords with the wave function ELCW.    The dashed curve ELS represent the lower bound given by the sum approximation. Accurate numerical data (dotted-curve) EX is shown for comparison.

% -------------------------------------------------------------------------
\np
\hbox{\vbox{\psfig{figure=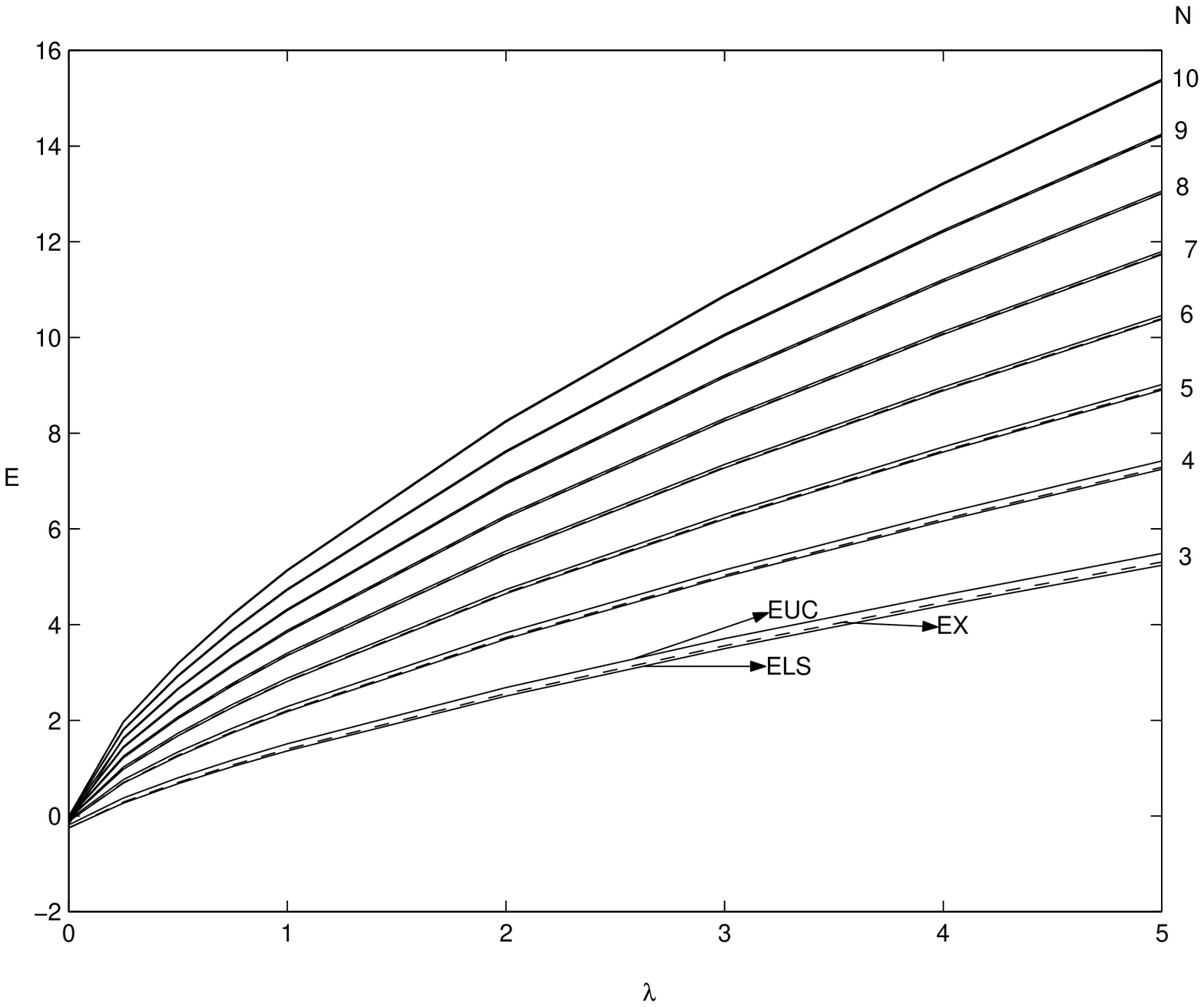,height=4in,width=5in,silent=}}}
\title{Figure 5.}
\nl Bounds on the eigenvalues $E_{1 0}^{N}(\lambda)$ corresponding to the Coulomb-plus-linear potential $V(r) = -1/r + \lambda r$ in $N$ dimensions. Upper bounds EUC by the generalized comparison theorem, lower bounds ELS by the sum approximation, and accurate numerical data (dashed-line), for $n=1,$ $\ell = 0,$ and $N=3,4,\dots 7.$ By Theorem~2 we know that the same curves apply also to $\ell > 0$ since $E_{1\ell}^{N} = E_{1 0}^{N+2\ell}.$     

% ------------------------------------------------------------------------- 
\end